# Optically reconfigurable polarized emission in Germanium

Sebastiano De Cesari*, Roberto Bergamaschini, Elisa Vitiello, Anna Giorgioni, and Fabio Pezzoli*

LNESS and Dipartimento di Scienza dei Materiali, Università di Milano-Bicocca, via Cozzi 55, 20125 Milano, Italy

Corresponding Authors: fabio.pezzoli@unimib.it; sebastiano.decesari@unimib.it

**Light polarization can conveniently encode information. Yet, the ability to steer polarized optical fields is notably demanding but crucial to develop practical methods for data encryption and to gather fundamental insights into light-matter interactions. Here we demonstrate the dynamic manipulation of the chirality of light in the telecom wavelength regime. This unique possibility is enrooted in the multivalley nature of the conduction band of a conventional semiconductor, namely Ge. In particular, we demonstrate that optical pumping suffices to govern the kinetics of spin-polarized carriers and eventually the chirality of the radiative recombination. We found that the polarization of the optical transitions can be remarkably swept through orthogonal eigenstates without magnetic field control or phase shifter coupling. Our results anticipate spin-enhanced optoelectronic functionalities in materials backed by the mainstream microelectronics industry and provide guiding information for the selection and design of functionalized emitters based on novel group IV compounds.**

The ever-increasing demand for fast, secure data processing and transmission has sparked an intense search for high-volume and cost-effective approaches to blend sensing [1], storage [2, 3] and communication [4, 5] functionalities within semiconductor devices. Spin and photon degrees of freedom have emerged as competing candidates in the quest to extend or even substitute current electronics based on charge transport [6-8]. Their entanglement, albeit challenging, holds the promise of a greatest payoff yielding prospects for a new family of multifunctional, high-performance and energy-efficient devices like spin-based lasers and LEDs [9-12]. Merging magnetic and photonic functionalities would open the way to a wide array of advanced operations encompassing (i) large-bandwidth and fast signal switching by polarization multiplexing in data transmission protocols [13-15], (ii) optical writing of nanomagnets for high-density and high-speed memories [16-18], (iii) encoding solid-state quantum bits on flying photons for chip-scale and long haul optical fibre communication [19-21], and (iv) lab-on-chip biological sensing of chiral-dependent structures and chemical phenomena [22-24]. The coupling between the angular momentum of photons and electrons via optical selection rules was indeed exploited in the pioneering demonstration of ultrafast optical switching through coherent spin rotations [25], and anticipated lasing dynamics exceeding 100 GHz [25-27]. Although this essentially applies to devices working at cryogenic temperature, relying on III-V compounds and magnetic fields as large as few Tesla, it discloses the true potential of spin-optoelectronics to outperform its conventional counterpart [25- 29].

The joint implementation of spintronics and photonics within the same Si platform offers device concepts with more realistic and scalable application perspectives owning to the mature and vast infrastructure associated with the mainstream microelectronic industry [30]. The core component of such hybrid spin-optoelectronic architectures would be a light source that fulfils two major requirements: The compatibility with Si-based complementary metal-oxide-semiconductor (CMOS) processing and the capability of spontaneous or stimulated emission of photons with on-demand and well-defined states of polarization. On one hand, the need of a suitable CMOS-compatible light source has been subject of growing interest within the group IV community and only very recently has led to the seminal demonstrations of lasing action in tensile-strained [31, 32] as well as Sn-alloyed germanium [33, 34]. Those reports, although limited by low efficiency and low temperature operation, are important cornerstones for the implementation of active photonic components. Switching the state of light polarization, however, turns out to be a major hurdle. The lack of simple turning knobs is yet a crucial roadblock as current approaches mostly rely on strong external magnetic fields or on bulky phase shifters that typically require slow mechanical movements [11, 25]. The available techniques, however, do not allow a compact and monolithic design of the kind needed for integrated spin-optoelectronics architectures.

In this work, we overcome such an obstacle demonstrating all-optical control of chiral light emission in group IV materials. We propose a simple scheme based on engineered spin-dynamics. This allows the manipulation of the spin polarization of selected minima or valleys of a semiconductor, and we prove this approach to be a powerful means to achieve the on-demand definition of the angular momentum of the emitted light. Here we validate this concept by applying Stokes polarimetry to the low temperature direct gap photoluminescence (PL) of Ge [35], whose multi-valleyed structure of the conduction band (CB) naturally suites best this purpose [36], and whose optical properties match well a desired operation at telecommunication wavelengths [37].

In our experiments, the polarization of the direct gap PL was found to be continuously tunable between the left- ($|L\rangle$) and right-handed ($|R\rangle$) circular polarization eigenstates. Notably, this was achieved by optical pumping at a fixed excitation energy and by solely modulating the laser excitation fluency. Moreover, we show that linearly polarized PL, i.e. ½ ($|L\rangle+|R\rangle$), can be obtained as a consequence of a coherent superposition of two equal populations of electrons having opposite spin orientations.

## Results

**Concept of dynamical spin control.** Before applying chiroptical spectroscopy, we illustrate the intriguing spin dynamics that pertains to the multivalley CB of Ge and elaborate on the proposal of its possible exploitation to attain an unmatched polarization control.

In Ge, the excitation via infrared circularly polarized light with energy almost resonant with the direct gap threshold results in optical spin orientation (see Fig. 1a-b) [35, 38]. Vertical transitions in the vicinity of the $\Gamma$ point at the centre of the momentum space create a non-equilibrium population of spin-polarized electron and hole pairs characterized by the quantum numbers $|J, J_z\rangle$ describing the total angular momentum, $J$, and its projection, $J_z$, along the quantization axis. Under our experimental conditions, light impinges perpendicularly onto the sample surface and the spins point towards the $z$ direction defined by the photon momentum (see Fig. 1c).

By virtue of spin orbit coupling and of the dipole-allowed selection rules, two subsets of electrons can be notably initialized in the CB with different kinetic energies and with spins pointing in opposite directions (Fig. 1b). In particular, upon $|R\rangle$ excitation the largest fraction of electrons is optically coupled to the upper VB states, gaining a sizeable momentum and, on average, a spin-down direction, i.e. $|1/2, -1/2\rangle$ [6, 39]. Simultaneously, a minute amount of $|1/2, 1/2\rangle$ spin-up electrons can be promoted at the edge of the $\Gamma$ valley via transitions involving split-off VB states (Fig. 1a-b) [6, 39, 40].

In the process of reaching thermal equilibrium with the crystal, the photogenerated holes relax towards the top of the valence band (VB) at $\Gamma$, while electrons sample the minima landscape of the CB. The latter results in puzzling spin-dependent phenomena and represents a remarkable difference with respect to the well-known case of direct gap materials like III-V compounds, where $|1/2, \pm 1/2\rangle$ electrons can simply thermalize at $\Gamma$ [35, 36].

In indirect gap materials, electrons ultimately scatter out of the optically coupled Γ region, and are eventually collected at the global minima. In Ge these are energy-degenerate and located along the four equivalent [111] directions at the *L* point of the Brillouin zone as shown in Fig 1a.

It is worth noting that in Ge the additional satellite *X* valleys, owning to a large density of states, offer a preferential kinetic pathway for the spin-down electrons generated above the edge of the Γ valley. Literature works pointed out the importance of the energy relaxation channel opened up by the *X* valleys [41, 42] and the Coulomb interactions occurring therein [35]. In highly-doped samples, indeed, collective and binary collision processes with the Fermi-Dirac distributed background carriers have been argued to particularly enhance the occurrence of scattering events guiding a minute fraction of spin-down electrons back to the original Γ valley [35, 43]. This mechanism, in turn, is meant to cause a reversal of the direct gap circular polarization with respect to the case of weakly-doped samples, whose band-edge luminescence is determined by the pristine lower-energy spin-up electrons [35]. It should be noted that spin-orbit coupling mixes the spin-dependent wavefunctions of VB Bloch states, typically resulting in an unpolarized hole ensemble [44]. On the contrary, the spin relaxation time of electrons greatly exceeds the lifetime at Γ and thus dictates the state of the direct gap PL polarization [36, 45].

Inspired by these findings, we can explore the unique possibility to utilize such an effective momentum-spin locking of CB electrons to actively manipulate the dwelling of a selected spin ensemble at the zone centre, thereby altering the spin imbalance accumulated at the direct valley through the initial optical orientation process. In the following, we leverage photo-injection to intentionally modify the free-carrier concentration. By doing so, we demonstrate how to optically manoeuvre Coulomb-mediated interactions and achieve the dynamical control over the PL polarization.

**Stokes polarimetry applied to Ge**. We studied Ge:As wafers whose doping concentration is $8.3 \times 10^{16}$ cm$^{-3}$. As it will be discussed in the following, the choice of the doping level and lattice temperature facilitates the successful control of the polarization.

Figure 2a shows the direct gap PL under $|R\rangle$ excitation at 1.165 eV. All the measurements have been carried out at 4K. The PL demonstrates a peak at ~0.87 eV. The skewed lineshape stems from the joint density of states and the Maxwell-Boltzmann-like distribution of the carriers within their bands. The former determines the low energy threshold of the PL feature, while the latter yields an exponential-like tail on the high-energy side of the peak.

Figure 2a shows specifically the power-dependent PL characteristics. Besides the expected decrease of the peak amplitude with the number of photoinjected carriers, the PL demonstrates a puzzling polarization pattern. Figure 2b reports a color-coded map of the normalized PL peak intensity as a function of the angle of a polarization analyser, namely a quarter wave plate followed by a linear polarizer. Figure 2b shows the data gathered for all the values of the incident power densities accessible in our experiment. Three well-defined regimes can be observed. At low pump power, the pattern reveals a sinusoidal behaviour with two maxima (minima) at 45° (135°) and 225° (315°). Such modulation unambiguously unveils a $|L\rangle$ polarization eigenstate, which is cross-polarized with respect to the $|R\rangle$ excitation [46, 47]. In the high injection regime, however, the periodic feature is surprisingly out-of-phase with respect to the low power regime. Indeed, the light field turns out to be in the orthogonal $|R\rangle$ eigenstate, as demonstrated by the two maxima (minima) π-shifted at 135° (45°) and 315° (225°) [46, 47]. Finally, at intermediate densities, the modulation frequency of the PL intensity oscillation demonstrates a twofold increase, heralding unambiguously a linearly polarized emission [46]. From the modulation pattern of Fig. 2b, we derived the Stokes vectors and summarized the associated polarized component of the light field on the Poincaré sphere of Fig. 2c [46, 47]. Stokes polarimetry clarifies the effectiveness of the incident power density, i.e. carrier population, in steering the polarized component of the PL along the sphere. Remarkably, a full helicity inversion of the direct gap PL can be attained, eventually swapping the pure orthogonal $|R\rangle$ and $|L\rangle$ states.

Such subtle finding, which notably do not occur in intrinsic or heavily doped samples (see Supplementary Fig. 1), can be naturally understood in the framework of the electron spin dynamics taking place in a multivalley material, as described in the previous section. The larger the density of the photo-generated carriers, the larger the probability that electrons experience intravalley Coulomb-mediated collisions during the thermalization process.

This finding discloses a new approach for designing polarized emitters, and opens up potential applications such as novel integrated polarized photon sources and spin-based lasers.

**Kinetic model**. In the following, we elaborate further on the key role of the free carrier density and Coulomb interactions in dictating the energy relaxation and eventually the PL polarization. To this purpose, we model the cooling processes experienced by hot electrons in the CB extending the three-state model proposed by Stanton and Bailey [48]. In particular, we explicitly include spin-dependent phenomena pertaining to a multi-valleyed material.

The simplest system that we can consider is schematically shown in the inset of Fig. 3. It includes an energy level near the edge of the $\Gamma$ valley, termed $\Gamma_<$, which is partly occupied by spin-up thermal electrons originated from the split-off VB. We assume that under steady state conditions, electrons can recombine radiatively through direct band gap transitions only via this state. A second higher energy level, $\Gamma_>$, hosts electrons optically coupled to the topmost VB states having a net spin down orientation. Finally, we consider a third level for the electrons being scattered to the $X$ valleys. It should be noted that, due to the indirect nature of the Ge band structure, the absolute $L$-valley minima of the CB are included in the model as a sink in which carriers can reside without contributing to the final polarization of the direct gap PL. The dynamics of the most relevant processes involving these states can be generalized as follows:

$$\begin{bmatrix} \dot{n}_{\Gamma_>^\downarrow} \\ \dot{n}_X \\ \dot{n}_{\Gamma_<^\downarrow} \\ \dot{n}_{\Gamma_<^\uparrow} \end{bmatrix} = \begin{bmatrix} -(\gamma_{\Gamma X} + \gamma_{\Gamma_> L}) & 0 & 0 & 0 \\ \gamma_{\Gamma X} & -(\gamma_{X\Gamma} + \gamma_{XL}) & 0 & 0 \\ 0 & \gamma_{X\Gamma} & -(R + \gamma_{\Gamma_< L}) & 0 \\ 0 & 0 & 0 & -(R + \gamma_{\Gamma_< L}) \end{bmatrix} \begin{bmatrix} n_{\Gamma_>^\downarrow} \\ n_X \\ n_{\Gamma_<^\downarrow} \\ n_{\Gamma_<^\uparrow} \end{bmatrix} + \begin{bmatrix} G^\downarrow \\ 0 \\ 0 \\ G^\uparrow \end{bmatrix} \quad (1)$$

where $n$ refers to the non-equilibrium carrier density induced by the optical excitation, $\gamma_{ij}$ is the net scattering rate from state $i$ to $j$, $R$ is the recombination rate, and $G^{\uparrow,\downarrow}$ are the generation rates for spin up ($\uparrow$) and spin down ($\downarrow$) electrons, respectively. Further details about the model and its parameters are given in the Supplementary Note 1.

After light absorption, the ultrafast phonon-mediated scattering acts on the spin-down carriers at $\Gamma_>$ populating the $X$ state. Coulomb-mediated interactions subsequently provide a very efficient energy loss channel [49, 50], eventually cooling the $X$-valley ensemble. The latter will then reach thermal equilibrium with the lattice by dwelling into the $L$ valleys. Notably, an increase of the pump power strengthen the electron-plasmon coupling, as in a material having effectively a larger doping level. This process substantially accelerates the hot carrier relaxation [50, 51] and, in turns, contributes to the occurrence of scattering events that can guide spin-down electrons towards the low-energy $\Gamma_<$ state [35].

As an educated guess, we can therefore apply a linear approximation and expand the density-dependent $X$-to-$\Gamma$ transition rate into a first order series, that is $\gamma_{X\Gamma} \approx \gamma_{X\Gamma}^0 + \gamma_{X\Gamma}^1 \cdot n_X$. As demonstrated in the Supplementary Note 1, where we derived an analytic expression for the populations of the $\Gamma_<$ level under the steady state approximation, the overall $X$-to-$\Gamma$ backward scattering is governed by the pump power density, $D$, according to a square-root power law. In light of this finding, it can be expected that an increase of the pump fluency triggers an accumulation of a spin-down subset at the $\Gamma_<$ state, eventually modifying the circular polarization degree ($P_c$) of the PL.

Figure 3 compares the $P_c(D)$ curve predicted by the kinetic model with the circular polarization degree, $P_C = \frac{I^{|R\rangle} - I^{|L\rangle}}{I^{|R\rangle} + I^{|L\rangle}}$, experimentally determined at various excitation power densities by measuring

the PL intensity I of the $|R\rangle$ and $|L\rangle$ components in two Ge:As samples having the same nominal doping content. The striking agreement disclosed by Fig. 3 is an open evidence of the emergence of Coulomb coupling, particularly electron-plasmon interactions. It also demonstrates that the modelling, despite being simple, captures the physics of the spin-dependent carrier dynamics.

Finally, we note that according to the theory, the pump power regime explored in this work results in $\gamma_{X\Gamma}^0 \gg \gamma_{X\Gamma}^1 \cdot n_X$ in intrinsic or weakly doped samples, whereas the opposite holds for heavily doped Ge. Such constraints impede the observation of a power-induced helicity change at these extremal doping regimes, in nice agreement with the results shown in the Supplementary Fig. 1.

**Spin and energy distribution of Γ-electrons**. Here we focus on the intermediate pump power regime in order to analyse in greater details the linearly polarized emission, and additionally test the physical picture emerged from the previous discussion.

Figure 4a shows the detailed color-coded map of the modulation intensity of the PL peak for a value of the incident power density of about 1.2 kW/cm². For convenience, the energy scale has been shifted with respect to the peak maximum with the aim to better emphasize the PL contributions of carries that differ for their kinetic energy. As previously discussed, a pump fluency of 1.2 kW/cm² leads to the observation of linearly polarized emission at the peak energy, i.e. ΔE=0 (see Fig. 2b and the middle panel of Fig. 4b).

A fine energy- and polarization-resolved analysis of the direct gap PL under this excitation condition unveils, however, a more elaborate and intriguing scenario. Indeed, a modest energy detuning of few tens of meV from the peak maximum remarkably demonstrates circularly polarized emission, see Fig. 4b and 4c. Chiefly, the high energy tail of the PL peak, i.e. ΔE>0, yields a right-handed ($|R\rangle$) circular polarization eigenstate, whilst the low energy side (ΔE<0) unambiguously discloses the pattern of a cross-polarized $|L\rangle$ eigenstate.

This evidence corroborates the modelling of the kinetics of the spin-polarized carriers. Above all, it clarifies that the linearly polarized photons emitted at the PL peak are due to a genuine superposition of the circular basis, namely ½ ($|L\rangle$+ $|R\rangle$), stemming from the spin-resolved energy spectrum of the $\Gamma$-valley electrons. Besides further strengthening the physical picture of the optically reconfigurable polarization of the direct gap PL, such a finding is a direct manifestation that a coherent spin dynamics sets in during the energy relaxation and recombination processes.

**Discussion**

We clarified the mechanisms that affect the energy relaxation of hot CB electrons in a technologically relevant material such as Ge. In particular, Coulomb interactions have been shown to reduce the ultrafast transfer rate towards the absolute minimum in the *L* valleys and enhance the backward scattering towards the zone centre. This permits to alter the relative weight of the two cross-polarized spin populations dwelling at $\Gamma$ and, in turn, to define the one that dominates the recombination events across the direct gap. We thus demonstrated that the kinetics of spin-polarized electrons in a multivalley conduction band material, such as Ge, can be readily modified by simply varying the free-carrier density. This can possibly open novel routes to address electron-plasmon coupling and hot charge-carrier interactions in a well-controlled solid-state environment, thereby providing relevant information for fields such as plasma physics and semiconductor plasmonics [52].

In view of the applications, the simple change of the intervalley scattering rate via the density of photogenerated carriers yields an effective modulation of the PL polarization without any external means, namely magnetic fields or optical retarders. To this final purpose, metamaterials [53-55], polariton condensates [56, 57], atomically thin transition metal dichalcogenides [58, 59] and quantum dots [60, 61] have been also put forward in the literature. Such approaches are however deficient, at a various degree, in scalability, integrability, in the spectral matching with the telecom range or, most notably, in the full helicity inversion. Instead, these are all expected benefits of our proposal that, we recognize, can be viable in applications that do not suffer strict requirements on photon losses.

The demonstration of tuneable polarization at the direct gap transition in Ge, albeit achieved at low temperatures, provides an important step towards the implementation of future spin-optoelectronics functionalities in a CMOS-compatible platform. Notably, theoretical investigations suggest that electrical spin injection can possibly occur at the Γ valley in Fe/MgO/Ge heterostructures [62], thus opening interesting perspectives for electrically controlled devices mimicking the optically-induced phenomena anticipated by this work.

Finally, our results point out the key role of the CB structure in dictating the spin dynamics and disclose a viable strategy for extending to other materials and device architectures the proposed concept of optically adjustable spin populations. It is worth noting that GeSn binary alloys have just been explored as direct gap materials that can be epitaxially grown on Si substrates. It can be thus expected that group IV compounds would soon lead to a wide and yet untapped range of multi-valleyed electronic structures, which can benefit from the guidance of our investigation to implement and optimize efficient and ultracompact polarized emitters.

**Methods**

**Experimental Setup**. PL measurements were carried out in a backscattering geometry using a closed-cycle cryostat. The working temperature was 4K and the incident power density was between 0.2 kW/cm$^2$ and 3.6 kW/cm$^2$. The samples were excited by a continuous wave Nd:YVO$_4$ laser operating at 1.165 eV. The incident light was right-handed circularly polarized $|R\rangle$, and the minimum laser spot diameter on the sample surface was measured to be (60±3) μm. The polarization state of emitted light was probed by means of a quarter-wave plate followed by a linear polarizer. The amplitude of the PL peak as a function of the analyser angle was measured by using a spectrometer equipped with an InGaAs array multiple-channel detector with a cut-off energy at about 0.755 eV. The energy accuracy of the whole system is of about 4 meV. Stokes parameters were provided by the analysis of peak amplitude modulation, allowing the characterization of the state of light polarization [46, 47].

**Acknowledgements**

We acknowledge M. Guzzi for his sustained support, E. Grilli and L. Miglio for fruitful discussions and M. Bonanomi for technical assistance. This work was supported by the Fondazione Cariplo through Grant SearchIV No. 2013-0623.

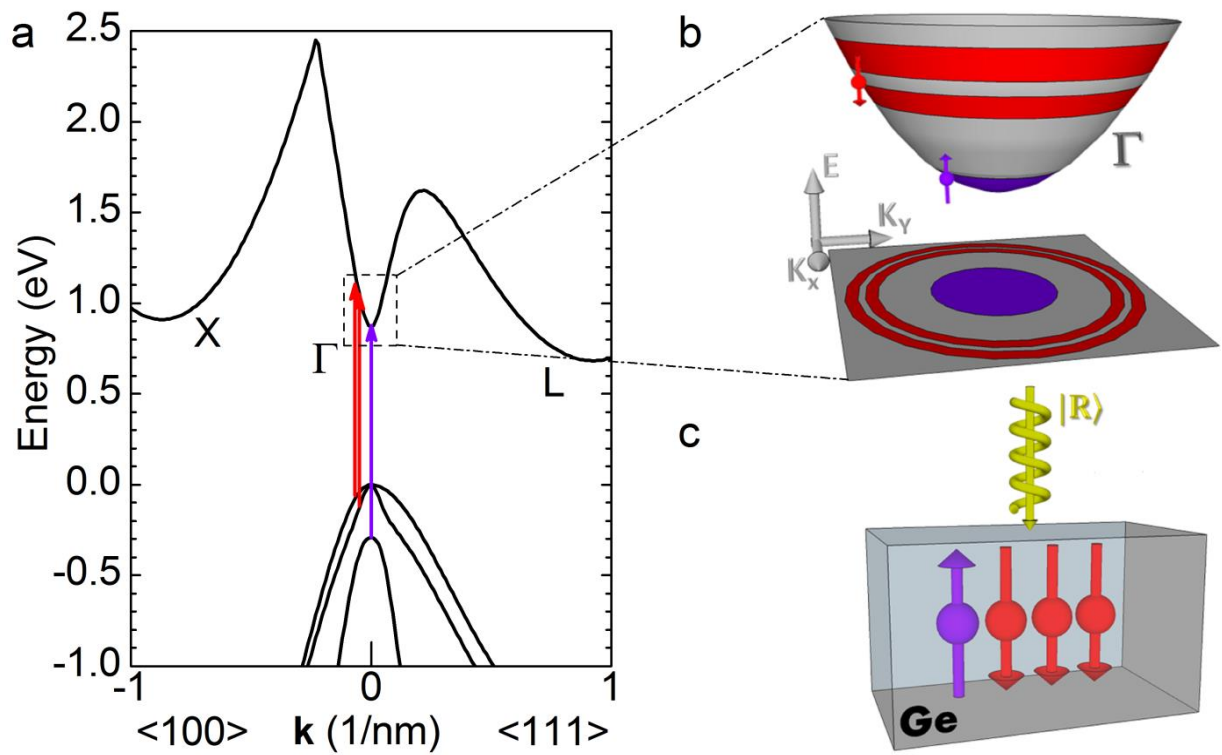

**Figure 1| Optical orientation process in Ge. a** Band structure of Ge along the [111] and [100] directions. **b** Schematic of the spin and momentum distribution of conduction band electrons located at the Γ valley. **c** Sketch of the geometry of the optical orientation experiment in Ge.

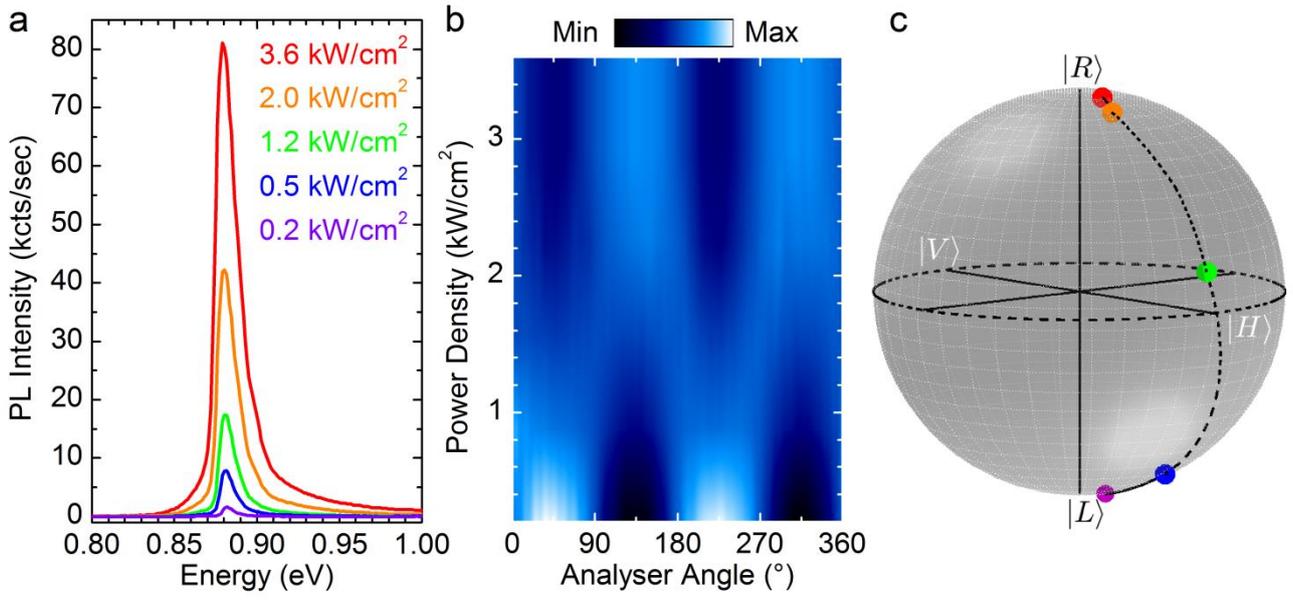

**Figure 2| Power-dependent PL characteristics and Stokes analysis. a** Direct gap photoluminescence (PL) spectra measured at 4K and under various excitation power densities in a n-type bulk Ge wafer; **b** Color-coded map of the PL peak intensity as a function of the angle of the polarization analyser for different pump power densities. **c** Poincaré sphere representation of the polarized PL components.

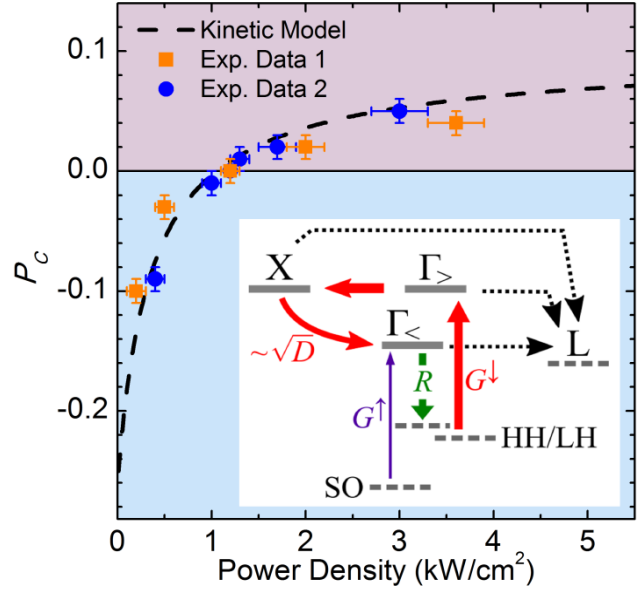

**Figure 3| Modelling of the power-dependent PL polarization.** Circular polarization degree of the direct gap PL measured as a function of the excitation power density in two n-doped Ge samples having the same doping content of $8.3\times10^{16}$ cm$^{-3}$ (red squares and blue circles). The black dashed line is the result of the kinetic model obtained by considering the excitation and recombination dynamics of spin polarized carriers schematically shown in the inset. Under circularly polarized laser excitation, spin up electrons ($G^\uparrow$) are photoexcited close to the Γ valley edge (Γ$_<$) from the split-off (SO) valence band state, while spin down electrons ($G^\downarrow$) are promoted from the heavy (HH) and light hole (LH) states to higher energies (Γ$_>$). The indirect band gap nature of Ge favours scattering of electrons out of the zone centre towards the *X* valleys and the absolute *L* minima. However, increasing the excitation power density (*D*), it is possible to strengthen the *X*-Γ backwards scattering, thus enhancing the spin down electron population contributing to the direct gap polarized PL (*R*). The different arrows thicknesses indicate the different efficiencies of the processes.

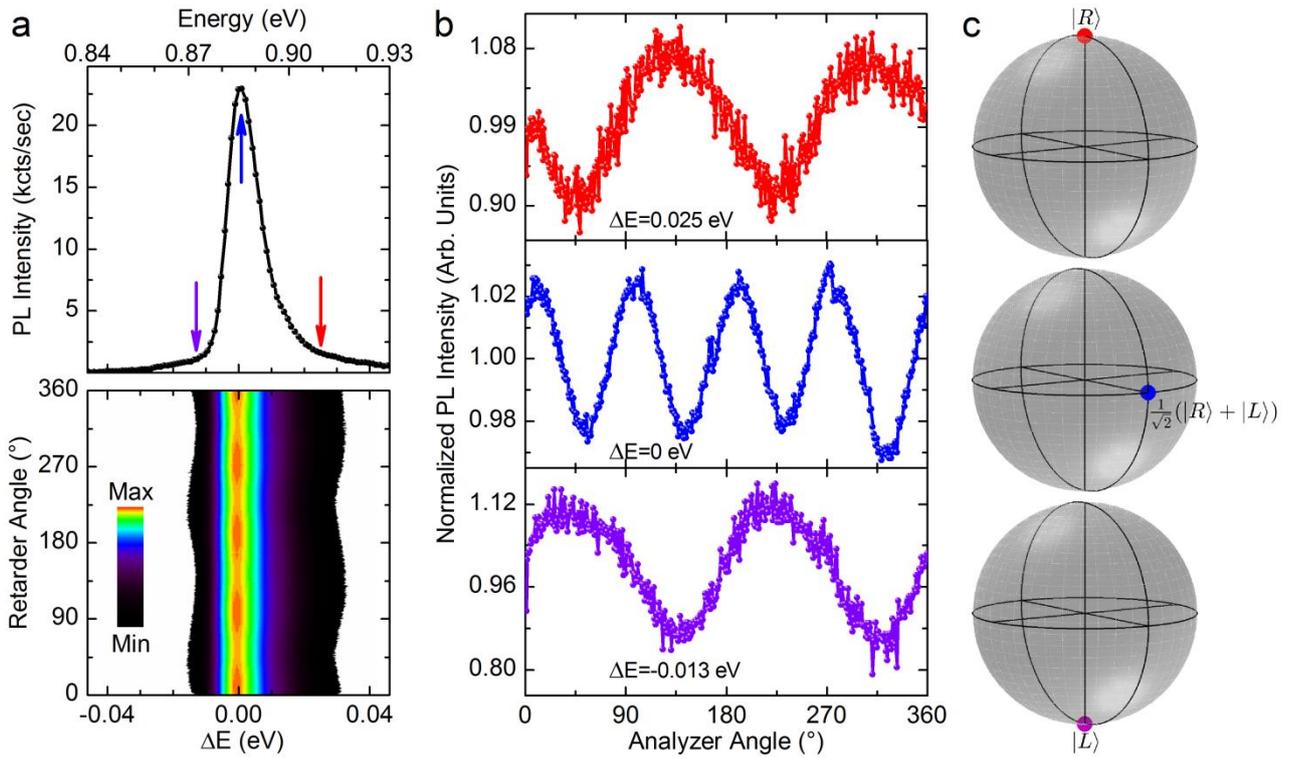

**Figure 4| Polarization-resolved energy spectrum of Ge. a.** Low-temperature direct gap PL of Ge under 1.2 kW/cm$^2$ excitation power density (upper panel). Color-coded map of the PL intensity describing the modulation of the direct-gap PL as a function of the analyser angle (lower panel). ΔE corresponds to the emission energy shifted with respect to the spectral position of the PL peak. **b.** Modulation of the PL peak intensity as a function of the angle of the polarization analyser. The data are spectrally resolved for photons emitted at the PL peak (cyan curve), at a positive (red curve) and at negative (violet curve) energy detuning ΔE. **c.** Representation on the Poincaré sphere of the polarized component of the PL corresponding to the ΔE values shown in b.